\documentclass[%
 aip,
 amsmath,amssymb,
 reprint,%
]{revtex4-1}
\usepackage{amsmath}
\usepackage{subfigure}
\usepackage{overpic}
\usepackage{siunitx}
\usepackage{float}
\usepackage{dcolumn}% Align table columns on decimal point
\usepackage{bm}% bold math

\makeatletter
\newcommand{\figcaption}{\def\@captype{figure}\caption}
\newcommand{\tabcaption}{\def\@captype{table}\caption}
\makeatother

\begin{document}

\preprint{AIP/123-QED}

\title{Spin image of an atomic vapor cell with a resolution smaller than the diffusion crosstalk free distance}
% Force line breaks with \\

\author{Hai-Feng Dong}
\email{hfdong@buaa.edu.cn.}
\affiliation{School of Instrumentation and Optoelectronic Engineering, Beihang University, Beijing 100083, China}
\affiliation{Quantum Sensing and Information Sensing Lab, Graduate School of China Academy of Engineering Physics, Beijing 100193, China}
\author{Jing-Ling Chen}
\affiliation{School of Instrumentation and Optoelectronic Engineering, Beihang University, Beijing 100083, China}
\author{Ji-Min Li}
\affiliation{School of Instrumentation and Optoelectronic Engineering, Beihang University, Beijing 100083, China}
\author{Chen Li}
\affiliation{School of Instrumentation and Optoelectronic Engineering, Beihang University, Beijing 100083, China}
\author{Ai-Xian Li}
\affiliation{Quantum Physics and Quantum Information Division, Beijing Computational Science Research Center, Beijing 100193, China}
\author{Nan Zhao}
\affiliation{Quantum Physics and Quantum Information Division, Beijing Computational Science Research Center, Beijing 100193, China}
\author{Fen-Zhuo Guo}
\affiliation{State Key Laboratory of Networking and Switching Technology, Beijing University of Posts and Telecommunications, Beijing 100876, China}
\affiliation{School of Science, Beijing University of Posts and Telecommunication, Beijing 100876, China}

\date{\today}% It is always \today, today,

\begin{abstract}
The diffusion crosstalk free distance is an important parameter for spin images in atomic vapor cells and is also regarded as a limit on the spatial resolution. However, by modulating the pumping light both spatially and temporally using a digital micromirror device, a spin image of a vapor cell has been obtained with a distinguishable stripe width of 13.7~$\mu$m, which is much smaller than the corresponding diffusion crosstalk free distance of $\sim$138~$\mu$m. The fundamental limit on the spatial resolution as determined by diffusion and the uncertainty principle is analyzed. 
\end{abstract}

\maketitle

\section{Introduction}

Measuring the spin spatial distribution in a vapor cell can help (i) optimize the spin-exchange optical pumping system, (ii) observe atomic diffusion directly, and (iii) characterize the dipolarization and desorption effects of the cell walls. It is also the basis of vapor cell magnetic field microscopy and can be used to image the inhomogeneous lightshift in the vapor cell, which is believed to be the reason of magnetic resonance signal asymmetry.\cite{wang2018inhomogeneous}

There are two main methods for measuring the spin spatial distribution in a vapor cell. One is optical magnetic resonance imaging (OMRI), which was pioneered in 1997 by Young et~al.\ \cite{young1997three} and Skalla et~al.\ \cite{skalla1997optical}. Both groups used a spatial field gradient to obtain the spin polarization distribution in a rubidium vapor cell. Young et~al.\ used a continuous-wave laser and a static field gradient, whereas Skalla et~al.\ used a pulsed laser and field gradient; the spatial resolutions of the two approaches were estimated to be 0.7~mm and 1~mm, respectively \cite{young1997three, skalla1997opticalPLA}. Although a stronger gradient is expected to improve the spatial resolution of OMRI, the effects of diffusion will counteract this improvement if the gradient is too large. Using the crossover field gradient given in Ref.~\cite{baranga1998alkali}, the spatial resolution defined by the linewidth of the point spread function is $\sqrt{D T_2}$, where $D$ is the diffusion coefficient and $1/T_2=R_p+R_{rel}$ is the sum of the pumping rate $R_p$ and the relaxation rate $R_{rel}$. Although Skalla et~al.\ define the diffusion-limited spatial resolution differently \cite{skalla1997opticalPLA}, the value corresponding to the crossover field gradient is also very close to that in Ref.~\cite{baranga1998alkali}, namely 0.9992$\sqrt{D T_2}$. Because the crossover gradient is usually much larger than any gradients actually used, $\sqrt{D T_2}$ is a practical limit on the spatial resolution of OMRI. In Ref.~\cite{savukov2015gradient}, a three-dimensional spin image was measured using OMRI with a spatial resolution of 0.8~mm $\times$ 1.2~mm $\times$ 1.4~mm. In Refs.~\cite{giel2000diffusion, weis2001motion}, OMRI and a mask with six 1-mm-diameter holes were used to detect the diffusion of atomic spins in a vapor cell.

The other method is multi-channel detection (MCD), in which the spin spatial distribution is measured using either a photodetector array or a charge-coupled device (CCD). MCD was used first by Romalis and colleagues in their spin-exchange relaxation-free (SERF) magnetometer with an adjacent channel spacing of 3~mm. Xia et~al.\ \cite{xia2006magnetoencephalography} used a 16$\times$16 photodetector array to measure the spin polarization distribution as a means of locating the source of magnetism in the brain. The diffusion crosstalk free distance $l=\sqrt{D T_2}$ was defined in Ref.~\cite{kim2014multi}, and the separation of the measurement channels is usually set to be smaller than this distance \cite{nishi2018high,zhang2019multi,mamishin2017novel}. Consequently, $\sqrt{D T_2}$ is also regarded as a limit on the spatial resolution of MCD. Rather than using a CCD, the combination of a digital micromirror device (DMD) and a photodetector is also used to detect the vapor-cell spin image caused by an alternating field \cite{Taue2017AC}.

In this letter, we use a DMD to modulate the pumping light both temporally and spatially to generate an equivalent magnetic field distribution with an adjacent channel spacing of 13.7~$\mu$m, and we obtain a spin image corresponding to this field distribution. The distinguishable stripe width is much smaller than the corresponding diffusion crosstalk free distance of 138~$\mu$m.  We analyze the fundamental limit on the spatial resolution as determined by diffusion and the uncertainty principle. With further improvement, the present method of spatial--temporal modulation may be used to characterize the fundamental spatial resolution of spin images in vapor cells.

\section{Experiment and Theoretical Analysis}
\label{sec:Experiment and Theoretical Analysis}

The experimental setup is shown in Fig.~\ref{fig:setup}. Light with a wavelength of 895~nm is generated by a distributed Bragg reflector (DBR) diode laser (PH895DBR240TS, Photodigm), whereupon the amplitude profile of the light is homogenized by a pinhole spatial optical filter (SOF). After being collimated by lens~L1, the light is elliptically polarized by a Glan--Taylor polarizer (P1) and a quarter-wave plate (QW1). A DMD (Vialux V-7001) with a pixel size of 13.7~$\mu$m $\times$ 13.7~$\mu$m is used to modulate the light both spatially and temporally. Lenses L2 and L3 are used to eliminate diffraction. To isolate the light completely when there is no polarization in the cell, another quarter-wave plate (QW2) and Glan--Taylor polarizer (P2) are set after the atomic vapor cell, which comprises cesium atoms in a 2.5~cm $\times$ 2.5~cm $\times$ 2.5~cm cubic cell made from quartz glass. The cell is filled with 600~torr of ${^4}$He buffer gas and 150~torr of N$_2$ quenching gas, placed in an oven made from boron nitride, and heated to 70$^\circ$C by applying a 20~kHz alternating current to twisted-pair wire. We use three-layer cylindrical magnetic shields with a shield factor of $\sim$10${^4}$. A set of coils inside the shields compensates for the residual magnetic field along $x$ and $y$ axes and generate a uniform field of 1,400~nT along the $y$ axis. 

\begin{center}
	\fbox{\includegraphics[width=85mm,height=41.38mm]{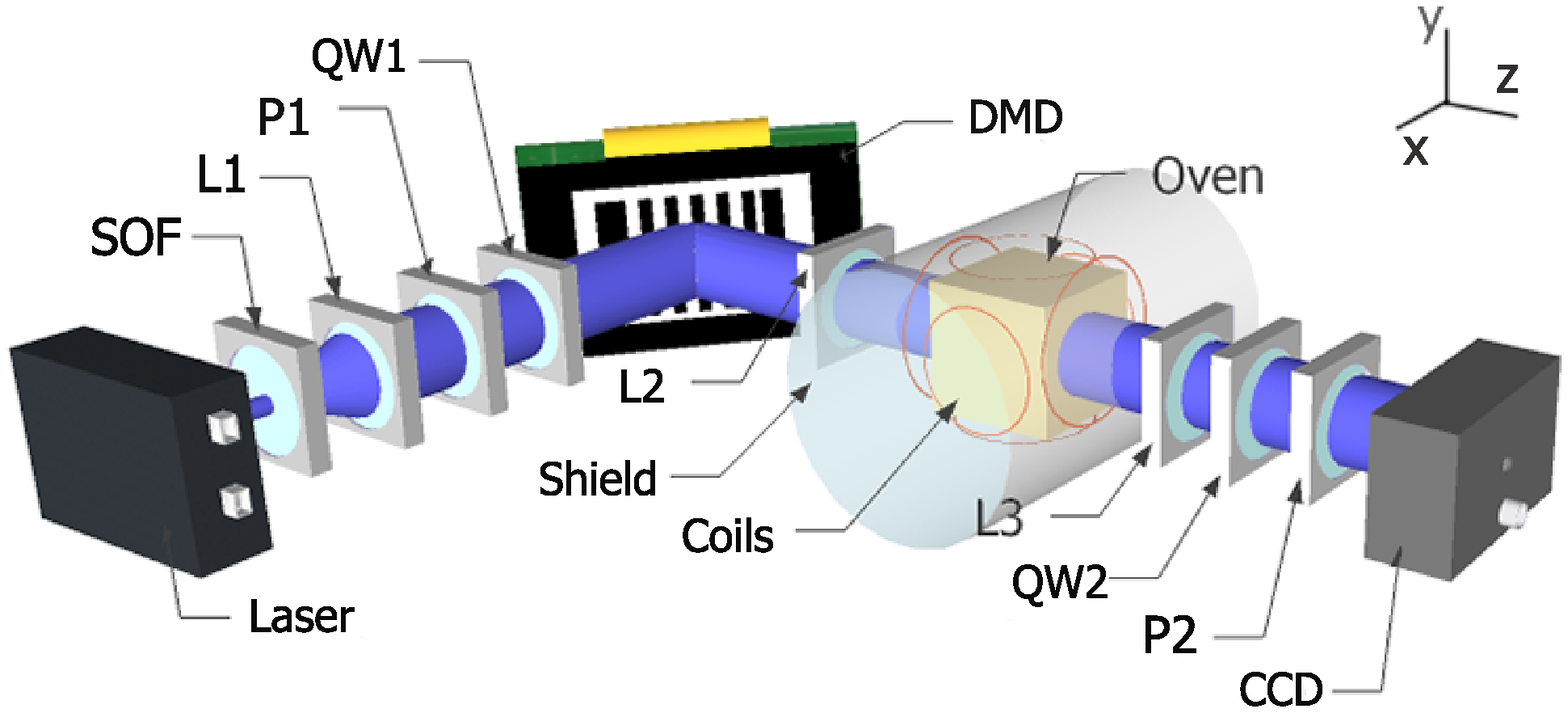}}
	\figcaption{Schematic of experimental setup (SOF: spatial optical filter; L1, L2, L3: lenses; P1, P2: polarizers; QW1, QW2: quarter-wave plates; DMD: digital micromirror device; CCD: charge-coupled device).}
	\label{fig:setup}
\end{center}

\begin{figure*}[htb]
	\centering 
	\subfigure[]{% 
		\includegraphics[width=50mm,height=53.3mm]{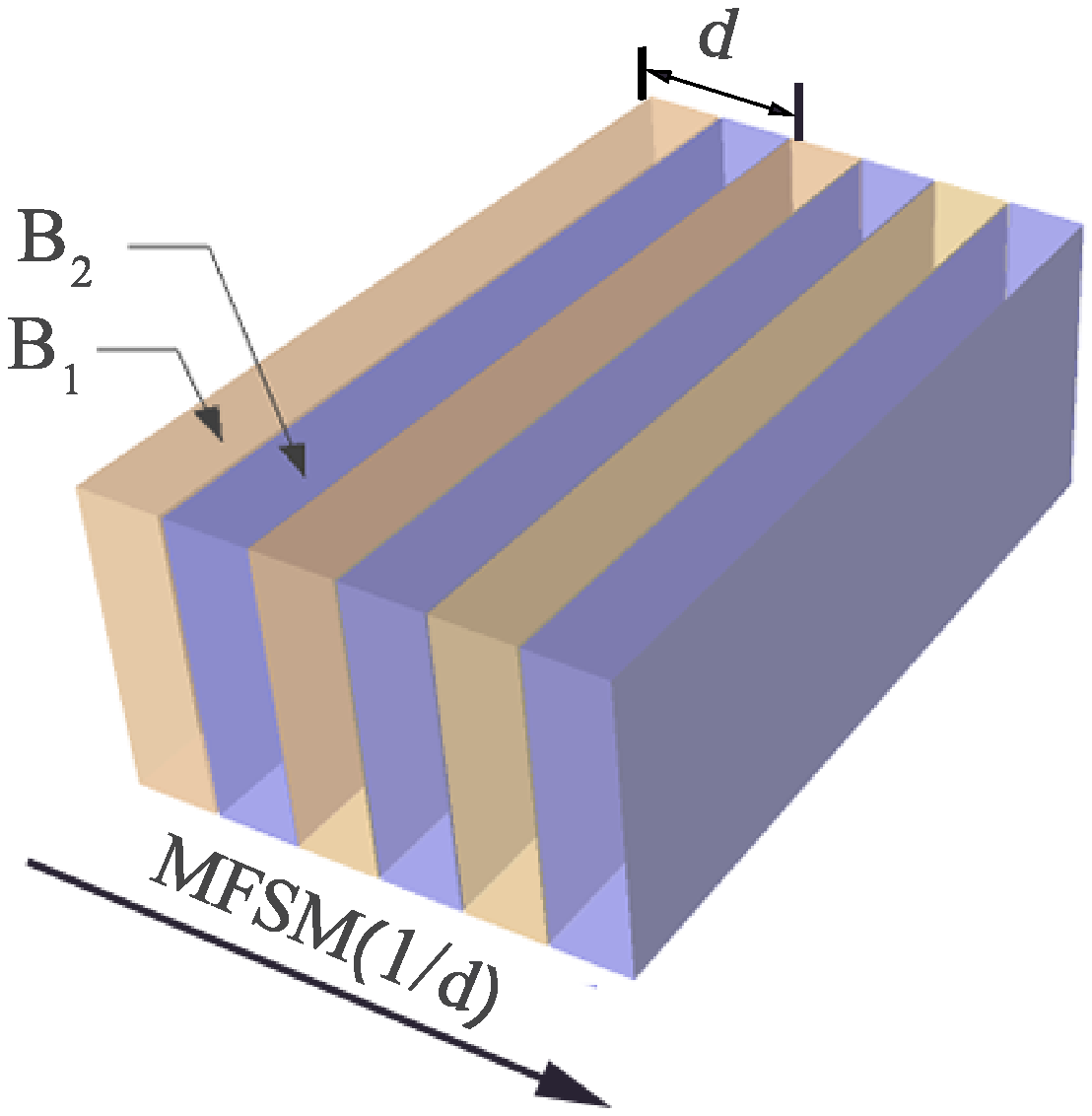}
		\label{fig:subfigure11}} 
	\quad
	\subfigure[]{% 
		\includegraphics[width=50mm,height=53.3mm]{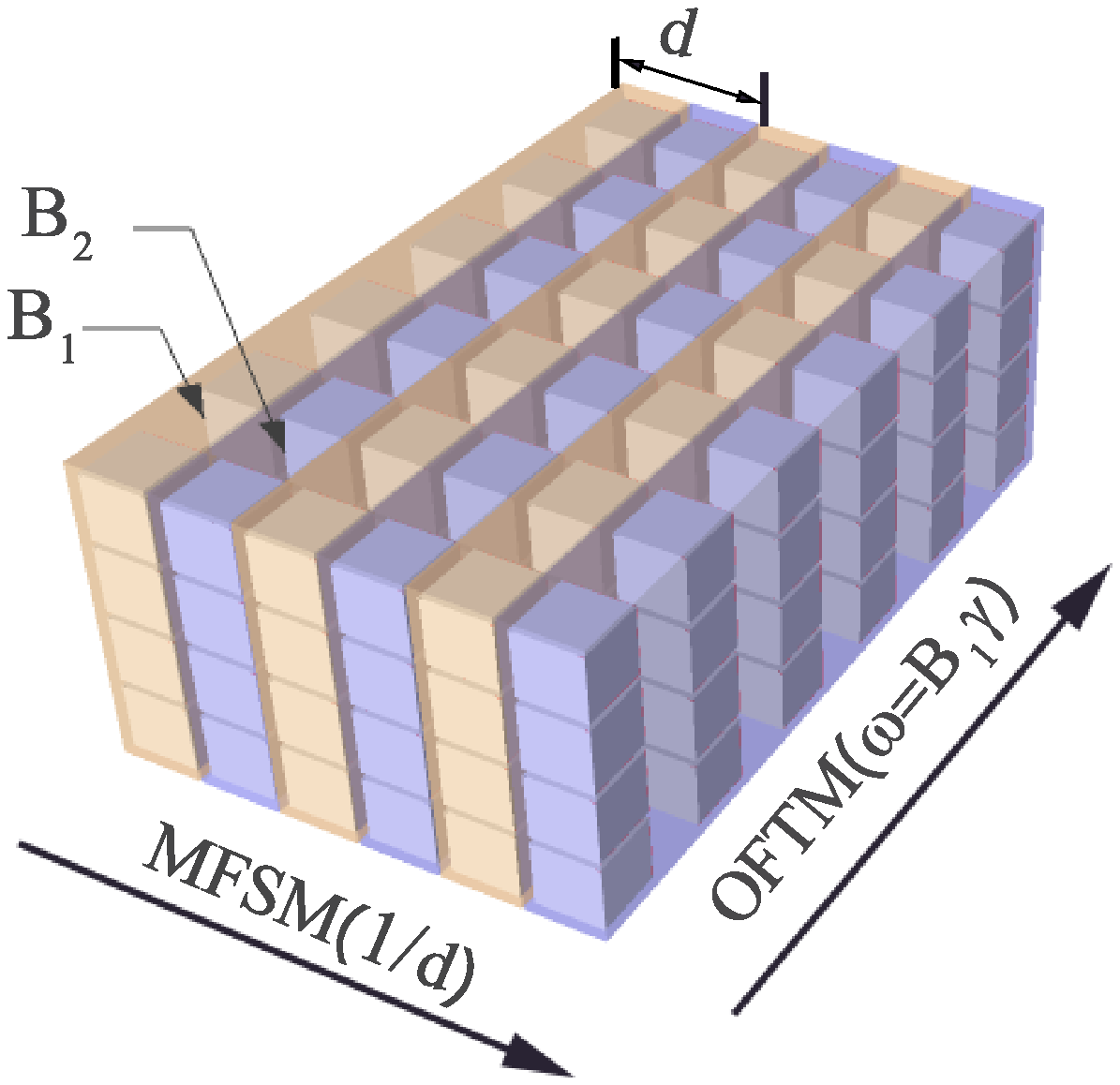}
		\label{fig:subfigure12}}
	\quad
	\subfigure[]{% 
		\includegraphics[width=50mm,height=53.3mm]{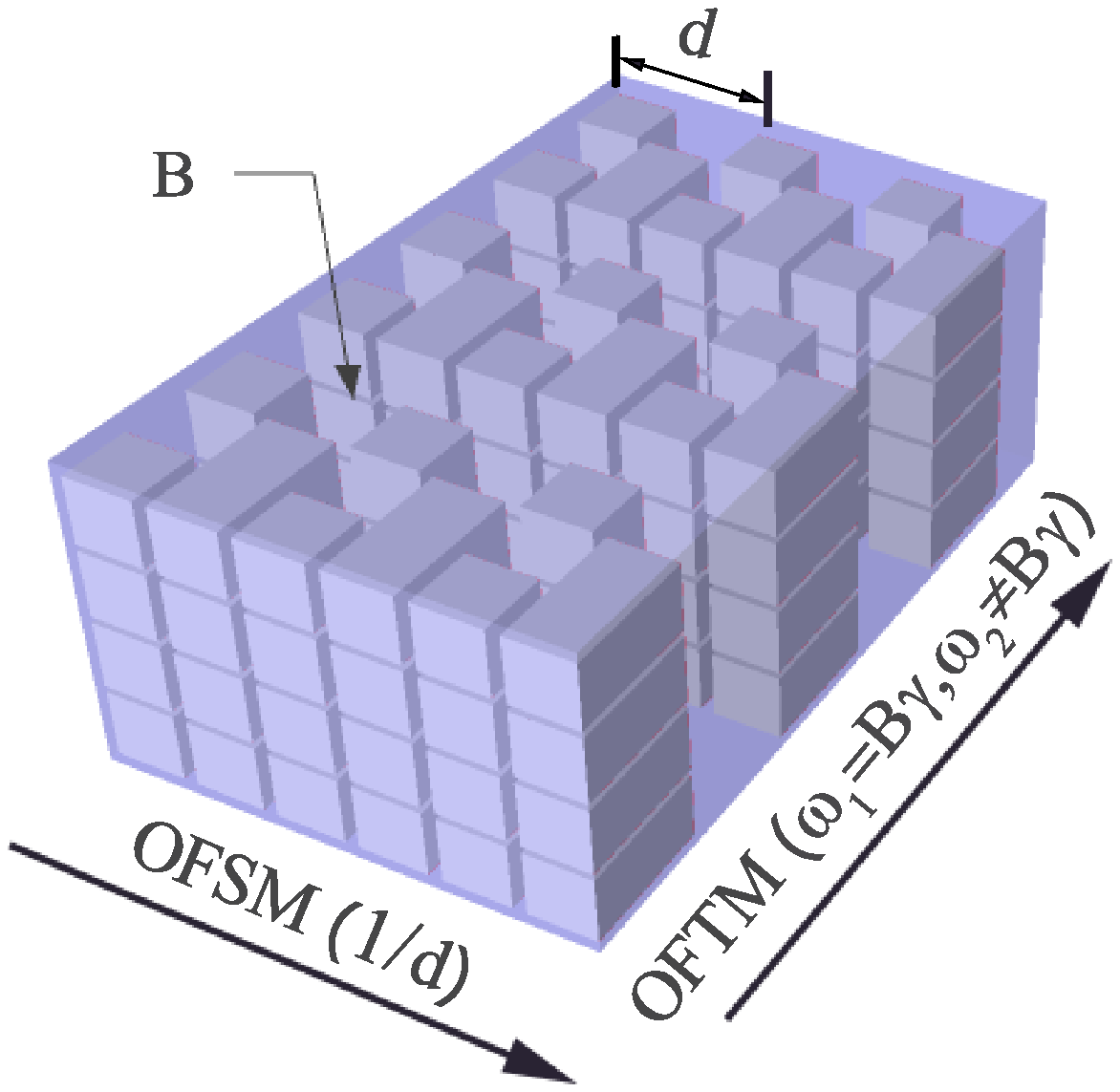}
		\label{fig:subfigure13}} 
	\caption{Illustration of generation of spin spatial distribution: (a) spatially-modulated magnetic field; (b) spatially-distributed synchronized pumping in field of (a); (c) spatially-distributed synchronized pumping in uniform field $B$. MFSM: magnetic-field spatial modulation; OFTM: optical-field temporal modulation; OFSM: optical-field spatial modulation; $d$: spatial period; $\gamma$: gyromagnetic ratio.} 
	\label{fig:figure} 
\end{figure*} 
With a spatially-modulated magnetic field shown in Fig.~\ref{fig:subfigure11}, where $B_1$ and $B_2$ alternate in space, one can generate a corresponding spin distribution. For example, one could use pulsed light with a temporal frequency of $\omega=B_1\gamma$ as shown in Fig.~\ref{fig:subfigure12}. In this way, the atomic spins in the regions with field $B_1$ are polarized because of the synchronized pumping, whereas those in the $B_2$ regions are less polarized or even unpolarized. Consequently, the modulated field pattern is transferred to the spin.

However, it is difficult to generate a spatially modulated magnetic field shown in Fig.~\ref{fig:subfigure11} using coils. We solve this problem by modulating the light both temporally and spatially using a DMD, as shown in Fig.~\ref{fig:subfigure13}. In this case, the magnetic field $B$ in the vapor cell is uniform, whereas the pulse frequency of the pumping light is modulated spatially. The atomic spins in the regions with $w_1=B\gamma$ are polarized synchronously, whereas those in the regions with $w_2\ne B\gamma$ are less polarized or even unpolarized. Consequently, the spin distribution generated in Fig.~\ref{fig:subfigure13} is the same as that in Fig.~\ref{fig:subfigure12}.

To reduce the background light and enhance the contrast of the spin image, we use two sets of quarter-wave plates and Glan--Taylor polarizers to realize elliptically-polarized pumping and orthogonal isolation, as shown in Fig.~\ref{fig:setup}.  In Ref.~\cite{shah2009spin}, elliptical pumping was used together with differential detection for a single-beam SERF magnetometer. Here, we combine elliptically-polarized pumping with orhtogonal isolation to overcome the difficulty of pixel matching in the image differential detection and to eliminate the blooming and smear effects of the CCD. After the first polarizer and quarter-wave plate, the light is polarized elliptically, comprising two circularly polarized components of opposite helicity with a Jones matrix as in 
\begin{equation}
\begin{aligned}
E & =\frac{E{_0}}{\sqrt{2}}\cos(\beta+\frac{\pi}{4})e^{-i\beta}\begin{bmatrix}
1     \\
i
\end{bmatrix}+\frac{E{_0}}{\sqrt{2}}\cos(\beta-\frac{\pi}{4})e^{i\beta}\begin{bmatrix}
1     \\
-i
\end{bmatrix},
\end{aligned}\label{eq:refname01}
\end{equation}
where $E_0$ is the light amplitude and $\beta$ is the angle between the optical axis of the first polarizer and that of the first wave plate. The first and second terms on the right-hand side of Eq.~(\ref{eq:refname01}) correspond to the left-circularly polarized light and the right-circularly polarized light, respectively. 

The atoms in the vapor cell are pumped by light of either polarization, giving rise to a spin polarization of   
\begin{equation}
P{_0}=\frac{R_{pr}-R_{pl}}{R_{pr}+R_{pl}+R_{rel}},
\label{eq:2}
\end{equation}
where $R_{pl}\propto(E{_0}\cos(\beta+\frac{\pi}{4}))^2$ and $R_{pr}\propto(E{_0}\cos(\frac{\beta-\pi}{4}))^2$ correspond to the left-circularly polarized light and the right-circularly polarized light, respectively. 

The optical rotation $\phi$ of the vapor cell is proportional to the spin polarization $P{_z}$ as
%$\frac{\Phi(\nu)\sigma(\nu)}{A}\propto$
\begin{equation}
\label{eq:opticalrotation}
\phi\approx{-\frac{1}{2}nc{r_e}lf_{D1}D1(\delta\upsilon)P_z},
\end{equation}
where $n$ is the atom density, $c$ is the speed of light, $r_e$ is the electron radius, $l$ is the length of the cell, $f_{D1}$ is the oscillation strength, and $D1 (\delta\upsilon)$ is the normalized absorption coefficient around the D1 line. 

The Jones matrix of the vapor cell is 
\begin{equation}
\label{eq:jones of cell}
{G_{cell}} =\begin{bmatrix}
\cos(\phi) & \sin(\phi)  \\
-\sin(\phi)  & \cos(\phi)
\end{bmatrix}.
\end{equation}

We aligned the optical axis of the second quarter-wave plate with the first one and set the optical axis of the second polarizer at an angle of $2\beta+\frac{\pi}{2}$ relative to that of the first polarizer. The output light amplitude after the second polarizer is 
\begin{equation}
\label{eq:3}
{E_{out}}={G_{P2}}{G_{QW2}}{G_{cell}}E,
\end{equation}
where ${G_{P2}}$ and ${G_{QW2}}$ are the Jones matrices of the second polarizer and the second quarter-wave plate, respectively, namely 
$$
{G_{P2}} =\begin{bmatrix}
sin^2(2\beta) & \frac{-sin(4\beta)}{2}   \\
\frac{-sin(4\beta)}{2} & cos^2(2\beta)
\end{bmatrix}, 
$$
$$
{G_{QW2}} =\begin{bmatrix}
cos^2(\beta)+isin^2(\beta) & \frac{sin(2\beta)}{2}-i\frac{sin(2\beta)}{2}    \\
\frac{sin(2\beta)}{2}-i\frac{sin(2\beta)}{2} & sin^2(\beta)+icos^2(\beta)
\end{bmatrix}. 
$$

\begin{figure}[htb]
	\centering 
	\includegraphics[width=70mm,height=42mm]{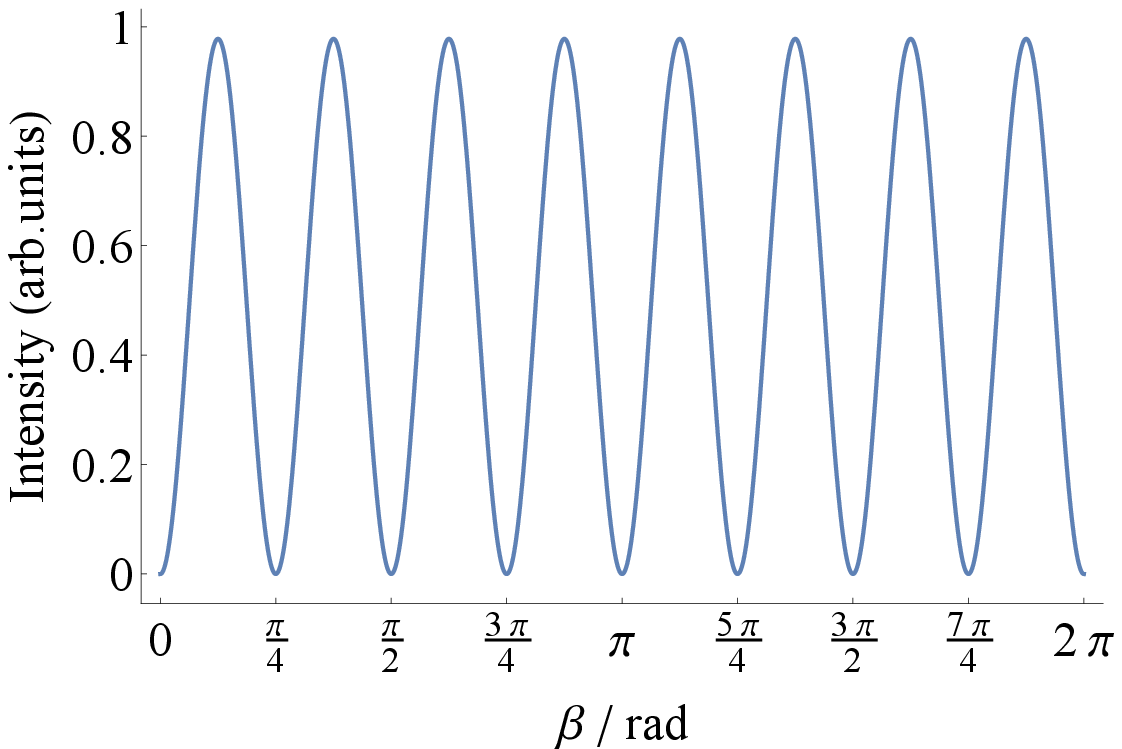}
	\caption{Output light intensity after second polarizer as a function of $\beta$.}
	\label{fig:I1} 
\end{figure}

%For certain vapor cell parameters, such as the temperature and buffer gas density, 
The relationship between the output light intensity $E_{out}^2$ and $\beta$ is shown in Fig.~\ref{fig:I1}. The intensity of the output light is zero at $\beta=0$ and $\frac{\pi}{4}$ and maximum at $\beta=\frac{\pi}{8}$. Consequently, we adjusted $\beta$ to be close to $\frac{\pi}{8}$ during the measurements.

\begin{figure}[h]
	\centering 
	\subfigure[]{% 
		\includegraphics[width=1.5in]{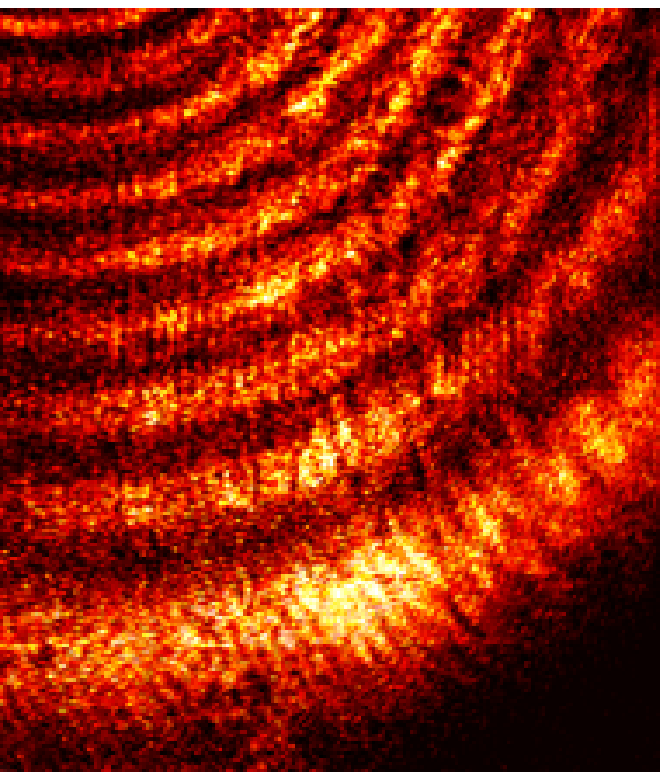}
		\label{fig:subfigure}} 
	\quad 
	\subfigure[]{% 
		\includegraphics[width=1.5in]{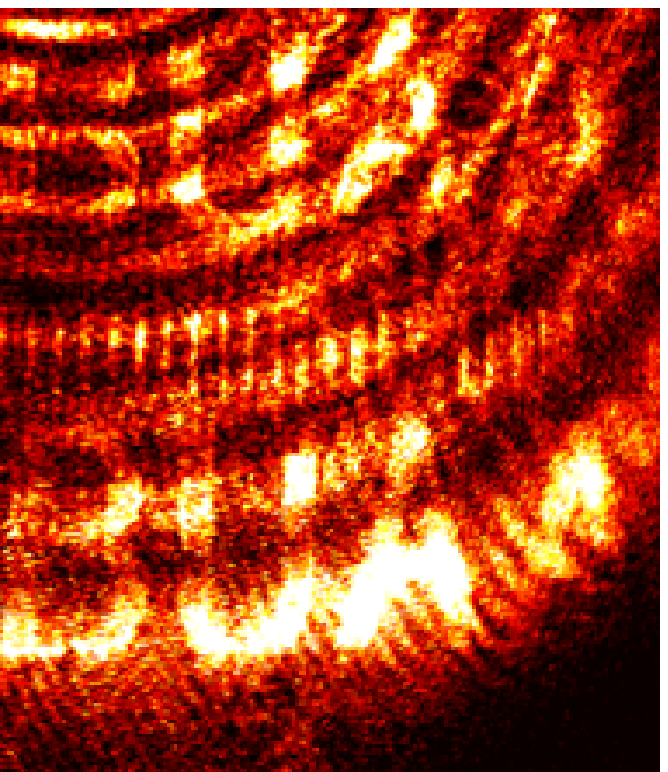}
		\label{fig:subfigure1}}
	\quad 
	\subfigure[]{% 
		\includegraphics[width=1.5in]{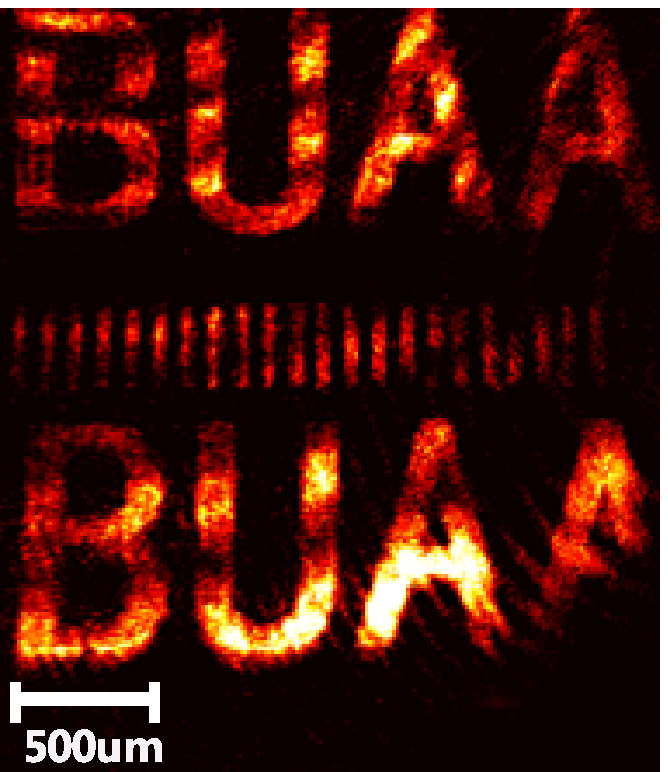}
		\label{fig:subfigure2}} 
	\quad 
	\subfigure[]{% 
		\includegraphics[width=1.5in]{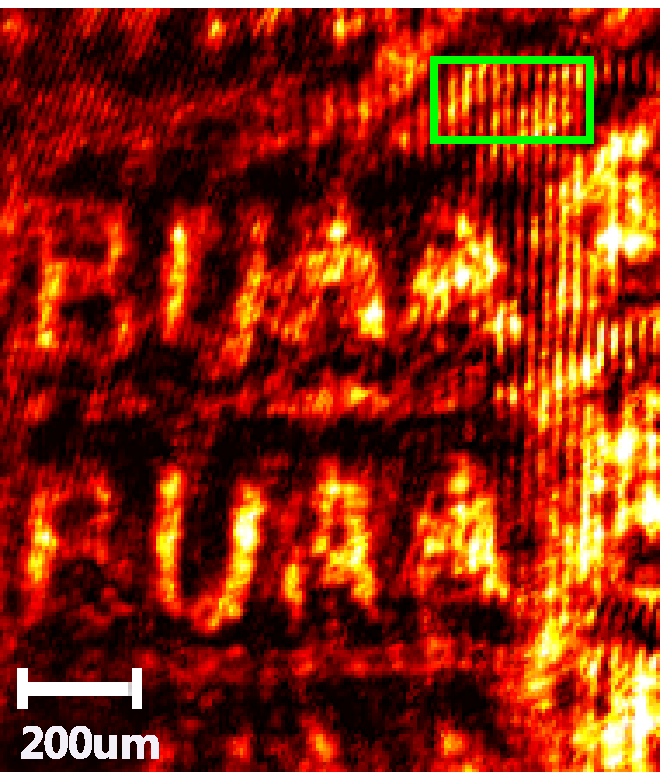} 
		\label{fig:subfigure3}} 
	\caption{Two-dimensional spin images of atomic vapor cell. (a) Background image without spin polarization. (b) Spin image written on background. (c) Spin image wiping off background by subtracting image (a) from image (b); the line width of the stripes in the middle of the figure corresponds to four column micromirrors with a width of 54.8~$\mu$m. (d) Another subtractive spin image, now showing a local linewidth of 13.7~$\mu$m (in green box).} 
	\label{fig:figure1} 
\end{figure} 

\section{Result and Discussion}

By modulating the light both temporally and spatially and using elliptically-polarized pumping and orthogonal isolation, we obtain two-dimensional spin images in the vapor cell as shown in Fig.~\ref{fig:figure1}. Figure~\ref{fig:subfigure} is the background image when there is no polarization anywhere in the vapor cell. Figure~\ref{fig:subfigure1} is the spin image written on the background by turning on the field $B=\omega_1\gamma$. The letters ``BUAA'' in the figure are illuminated because the spins in the corresponding regions are polarized synchronously. Figure~\ref{fig:subfigure2} is the spin image after wiping off the background by subtracting Fig.~\ref{fig:subfigure} from Fig.~\ref{fig:subfigure1}. The width of the stripes in the middle of Fig.~\ref{fig:subfigure2} is 54.8~$\mu$m and corresponds to four columns of DMD mirrors. In Fig.~\ref{fig:subfigure3}, a stripe width of 13.7~$\mu$m is shown locally in the green box, but the image of ``BUAA'' is no longer clear. A possible reason for this lack of clarity is light aberration, such as field curvature and stigmatism.

The transverse relaxation of the vapor cell is measured to be approximately 1.2~ms, while the diffusion coefficient is calculated to be 0.16 according to the temperature of the vapor cell and the pressures of N${_2}$ and ${^4}$He. With these two parameter values, the diffusion crosstalk distance $\sqrt{D T_2}$ is calculated to be approximately 138~$\mu$m, which is much larger than the distinguishable stripe width we obtain in the spin images. 

In what follows, we analyze the spatial resolution limited by diffusion and the uncertainty principle. Regarding diffusion, the polarization difference between adjacent regions will decrease with both the diffusion coefficient and the relaxation time. In Ref.~\cite{dong2019observation}, we analyzed the mean polarization of the synchronized pumping region, which is 
\begin{equation}
\overline {P_1} =P_0(1-\frac{2l_2}{l_1}\frac{(e^{\frac{d}{l_1}}-1)(e^{\frac{d}{l_2}}-1)}{dQ})\text{,}
\label{eq:6}
\end{equation}
where $l_{1}$ is the diffusion crosstalk distance without pumping light, $l_{2}$ is the diffusion crosstalk distance with pumping light, $d$ is the spatial period shown in Fig.~\ref{fig:figure} and $Q$ is a synthetic parameter. For $l_{1}$, $l_{2}$ and $Q$, we have 
$$l_{1}=\sqrt{\frac{D}{R_{rel}}},\quad  l_{2}=\sqrt{\frac{D}{R_{op}+R_{rel}}},$$
$$Q=\frac{1}{l_1}\left(e^{\frac{d}{l_1}}-1\right)\left(e^{\frac{d}{l_2}}+1\right)+\frac{1}{l_2}\left(e^{\frac{d}{l_1}}+1\right)\left(e^{\frac{d}{l_2}}-1\right).$$

Using the same method, one can obtain the mean polarization of the unsynchronized pumping region as 
\begin{equation}
\overline {P_2} =P_0(\frac{2l_1}{l_2}\frac{(e^{\frac{d}{l_1}}-1)(e^{\frac{d}{l_2}}-1)}{dQ})\text{,}
\label{eq:7}
\end{equation}
and the polarization difference between adjacent regions is 
\begin{equation}
\Delta P=\overline {P_1}-\overline {P_2}\text{.}
\label{eq:delta}
\end{equation} 

Regarding the uncertainty principle, the spin projection noise is related to the measurement volume as 
\begin{equation}
\label{polarizationuncertainty}
\delta{{P}}=2\sqrt{\frac{{{T}_{2}}}{qnAdt}},
\end{equation}
where $q$ is the slowing-down factor, $n$ is the atom density, $A$ is the cross-sectional area of the vapor cell, and $t$ is the measurement time. 

\begin{figure}[htbp]
	\centering 
	\includegraphics[width=70mm,height=42mm]{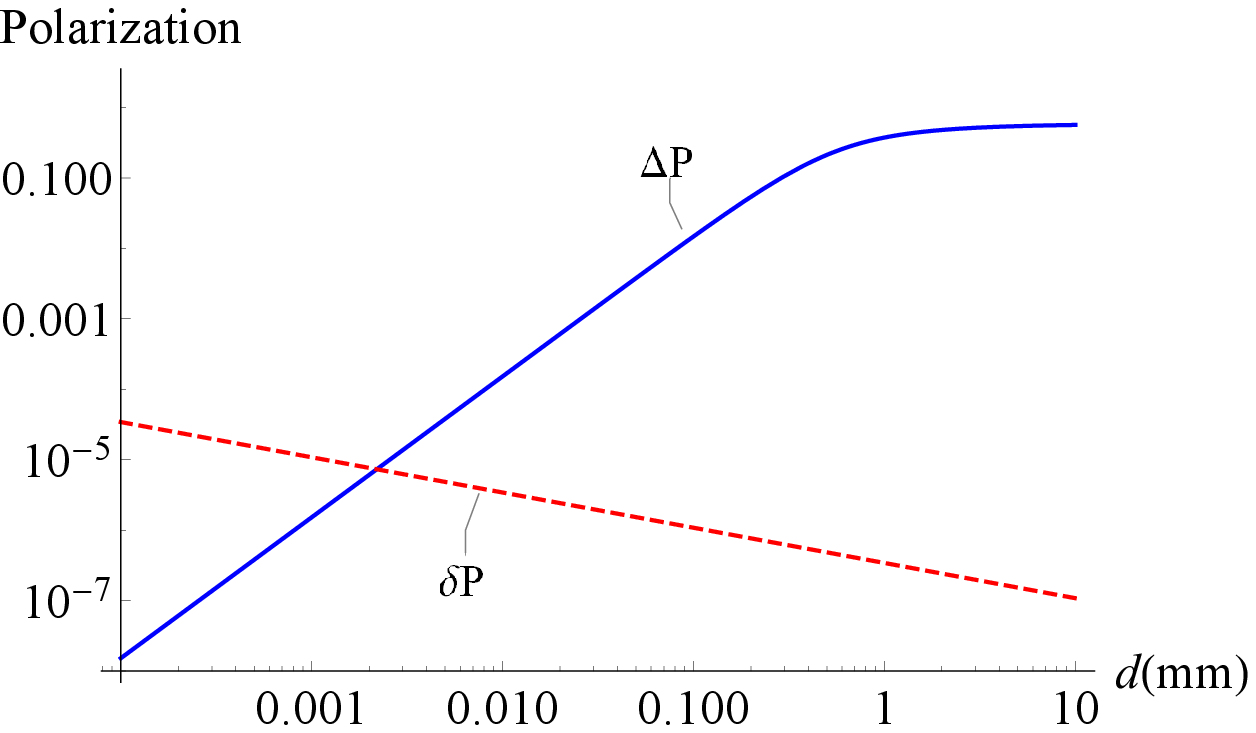}
	\caption{Theoretically calculated polarization difference between adjacent regions and corresponding spin projection noise.}
	\label{fig:I2} 
\end{figure} 

Figure~\ref{fig:I2} shows the results of theoretical calculations for the 2.5~cm $\times$ 2.5~cm $\times$ 2.5~cm vapor cell with typical parameters $D=0.16$~cm$^2$/s, $R_{op}=500$~s$^{-1}$, $R_{rel}=352$~s$^{-1}$, $n=10^{12}$~cm$^{-3}$, and $t=1$s. The position of the crossing point (at $d\approx2$~$\mu$m) represents the fundamental spatial resolution of the vapor cell determined by diffusion and the uncertainty principle. To the right of the crossing point, the spin polarization difference $\Delta P$ is larger than the spin projection noise $\delta P$, thus the adjacent region is distinguishable. To the left of the crossing point, $\Delta P$ is smaller than $\delta P$, thus the adjacent region is no longer distinguishable.

In a future experiment, the fundamental spatial resolution may be achievable by using (i) a high-resolution DMD and projection scaling to shrink the spatial modulation pattern and (ii) a vapor cell with higher buffer-gas pressure.

%Calculation shows that the fundamental spatial resolution of current atomic magneomter is from * to * $\mu$m corresponding to the range of D from * to * and the relaxation time from * to *.
\section*{Acknowledgments}
The authors thank Yanhua Wang, Feng Tang, Yi Zhen, Minjie Sun, Zhaohua Yang and Zhaohui Hu for helpful discussions and Uwe R{\"o}der for technical support. This work was supported by the National Natural Science Foundation of China (grant nos.\ 51675034 and 61273067) and the Natural Science Foundation of Beijing Municipality (grant no.\ 7172123). 

% Bibliography
\bibliography{spinimage}

\end{document}